# A Novel, Compact Optical Device for Estimating the Methane Emissions in Geological Environment

Sandipta Roy[a,b], Ramakrishnan Desikan[c,*], Siddhartha P. Duttagupta[b,d]

[a]*Centre for Nanotechnology and Science, Indian Institute of Technology Bombay, Mumbai-400076, India*
[b]*Centre of Excellence in Nanoelectronics, Indian Institute of Technology Bombay, Mumbai-400076, India*
[c]*Department of Earth Science, Indian Institute of Technology Bombay, Mumbai-400076, India*
[d]*Department of Electrical Engineering, Indian Institute of Technology Bombay, Mumbai-400076, India*

**Abstract**

Quantifying spontaneous, fugitive and venting related methane emissions are often difficult and cumbersome. However, auditing the methane emissions due to conventional and un-conventional hydrocarbon exploitation techniques are becoming necessary. Present generation compact chemical sensors are slower, degrade very fast, and are sensitive broad-spectrum gases. On the other hand, optical sensors are very fast in detection of gases and more precise and can be easily employed in various environments like boreholes and soils. In this study, we report development of an optical sensor that is methane specific, fast for real time applications and has tremendous application potential in the exploration of coal bed methane and other hydrocarbon reserves with methane as a major constituent. The detection process is based on the principle of spectroscopic absorption of light. The detector, NiSi Schottky diode, was fabricated and characterized exclusively for the 1.65 µm, narrow bandwidth methane absorption. The probe is of 20 cm length and comprises of a laser source and the NiSi detector aligned optically. This probe can be employed in boreholes, mine vents and soil layers for measuring real-time fluxes in methane concentrations. From the laboratory based experiments it is observed that the detection limits of the developed device is very low (3% by volume) and the response time of detection is very rapid (about 2 seconds). Based on the materials used, fabrication procedures adopted, sensitivity of the device and its compactness, the developed sensor can be considered as a novel, economic device for exploration of coal bed methane.

**Introduction:**

Seepage of thermogenic methane from the source rocks (eg. coal seams) is a common phenomenon in all sedimentary basins. Pathways for such migration often includes diffusion through pore spaces (in case of low permeable formations), and flows through high permeable formations and discontinuities (eg. faults, joints and cleavage planes). While, the rate of natural methane seepages is often slow, the induced methane seepages caused due to mining of hydrocarbons are fast and more hazardous. Common example of induced methane release is by coal mining wherein, a huge quantity of methane is released to the atmosphere through venting of open pit and underground operations. Fugitive release of methane from the conventional and non-conventional sources of fossil fuel exploitation is becoming another subject of global concern. It is estimated that methane emissions from the geological sources alone account for about 14% (40-80Tg/year) of current global values[1,2].

---

[*]email:- ramakrish@iitb.ac.in

The ever increasing awareness about the negative effects of methane emissions has been forcing the scientific community to audit the emission rates and evolve suitable techniques to minimize it. Consequently, many studies have been initiated globally to measure fugitive, venting and combustive emissions [1–3]. For this purpose, a wide range of direct (eg. soil gas analysis, bore hole measurements) and indirect (remote sensing) techniques have been employed over time and space [4–9].

In the literature, various types of methane detectors, such as non-dispersive infrared (NDIR) absorption spectroscopy [10–12], tunable diode laser absorption spectroscopy (TDLAS) [13,14], and catalytic combustion [15,16], metal-oxides [17] have been reported. However, most of the commercially available methane sensors are not sensitive to methane alone. Only optics-based sensors show selectivity and are able to detect to very low levels of concentration. In addition, optics-based sensors are based on the photonic absorption of the gas; therefore, these detectors respond almost instantaneously as gas comes through the pathway of the light. However, the cost of these sensors is comparatively very high, which prevents their extensive application. The increase in cost of such systems is generally attributed to the optical detector component. Thus, development of a cost effective infrared detector would enable wider application of optics-based sensors for gas detection. Schottky diodes are being considered as a cost-effective infrared detectors as they involve an inexpensive fabrication process [18]. However, this is often challenging due to difficulties in achieving the desired phase of the metal-semiconductor for detecting infrared and hence specific gas. Considering the explosion limit of methane (5% of methane concentration in air), the Schottky diode as methane detector has an excellent application potential for large scale deployment and precise monitoring of explosion vulnerability.

Methane absorption spectroscopy was performed and reported by several authors [19–22]. Methane has the strongest spectroscopic absorption in the infra-red (IR) region at 3.3 µm [23]. However, infrared sources at such a wavelength are expensive, thereby increasing the cost of the detection system. One of the solutions to this cost problem is to use the overtone absorption lines, which lie in the spectral region at 1.65 µm [24]. Although the absorption strength is low in this region, it has the advantage of not overlapping with water and other gaseous absorptions that can interfere with the methane response [22,25,26]. In addition, the infrared sources at that spectral region are also low-cost.

In this work, a methane specific, rapid detection system is demonstrated. This involves fabrication of a NiSi Schottky diode, its optical characterization, and demonstration as a methane detector. The detection system comprises a NiSi Schottky diode, a laser source, a gas chamber with flow controller and methane source. The proposed system is novel, cost effective and has the potential for extensive applications in detecting methane related explosion hazard.

# 1 Theory and Methodology

## 1.1 Methane absorption spectrum

Methane has tetrahedral symmetry and exhibits fundamental vibration modes at $v_1$ (2913 cm$^{-1}$ or 3.43 µm), $v_2$ (1533.3 cm$^{-1}$ or 6.52 µm), $v_3$ (3018.9 cm$^{-1}$ or 3.31 µm), and $v_4$ (1305.9 cm$^{-1}$ or 7.66 µm) [27,28]. All of these bands correspond to the far- and mid-infrared region. In the near-infrared (NIR) region, the overtones of the absorption band occur. The main overtones, which is a combination of rotational and vibrational modes of methane, consist of the $v_2 + 2v_3$ band (i.e., 1.35µm band) and the $2v_3$ band (i.e., 1.65 µm band). Due to the water and other gas absorption interferences at 1.35µm, the 1.65µm band is chosen in this study. A typical absorption spectrum of methane with the 1.65 µm absorption feature is shown in Figure *1*(a)[29].

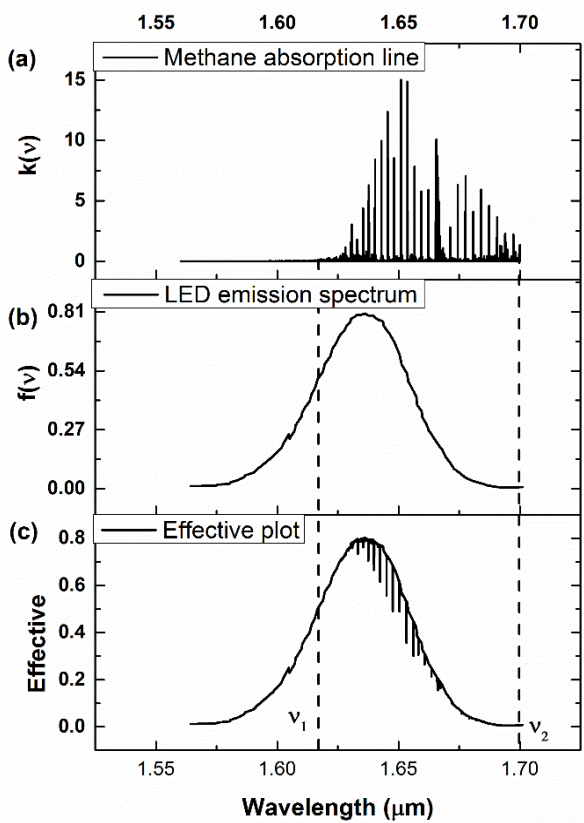

Figure 1: (a) The methane absorption line in the NIR region[29]; (b) emission spectrum of the LED; (c) overlap between the two spectra to define the band of detection.

## 1.2 Infrared Diode Emission spectra

The laser diode is selected in such a way that the emission band covers the maximum absorption region of methane at room temperature. In this experiment, a laser diode (part no: QSDM-1650-9 from Q-photonics [30]) of peak emission at 1.63 µm is used. The emission spectrum of the infrared diode laser is shown in Figure 1(b). Figure 1shows that the emission spectrum coincides with methane absorption. The laser beam has a divergence of 45° and 20° along the vertical and horizontal directions, respectively. To converge the beam on the sensor, a collimator (part no. C230TMD-C [31]) is used that has very low (0.5-1%) power loss due to reflection in the region of interest.

## 1.3 Detection principle

The methane molecule has a characteristic infrared absorption spectrum, as shown in Figure 1(a), which absorbs partially and transmits the rest of the irradiation. If the intensity of the LED is $P(v)$, then the emission distribution function is $f(v)$ at the wavelength $v$, with normalization constant K. The total intensity delivered by the diode can be expressed as

$$P_0 = \int P(v)dv = K \int f(v)dv \qquad (1)$$

If this LED output passes through the target gas cell of length $l$, in which the concentration of the target gas is c (volume percent), and assuming that all the light passes through the gas cell, then the output intensity from the cell is given by:

$$P_{Gas} = P_0 \exp(-k_{eff}cl) \qquad (2)$$

where $k_{eff}$ is the effective absorption cross-section of the gas due to the overlapping region of the gas absorption band and the LED emission band($v_1$ to $v_2$ shown in Figure 1(c)), which can be further expressed as:

$$k_{eff} = N_0 \int k(v) * f(v)dv;$$

where, $k(v)$ is the absorption cross-section at v, and $N_o$ (2.5×10$^{19}$ cm$^{-3}$) is the density of molecules per unit volume at normal atmospheric pressure and temperature.

Because the optical absorption of the gas is confined to the overlapping spectral region ($v_1$ to $v_2$), the effective absorption cross section becomes (the overlapping region is shown in Figure 1(c))

$$k_{eff} = N_0 \int_{v_1}^{v_2} k(v)\, dv$$

When the medium is ambient air, and assuming no absorption occurs in the medium, then $P_{Air} = P_0$. Therefore, equation 2 can be written as

$$P_{Gas} = P_{Air}\exp(-k_{eff}cl) \qquad (3)$$

Where $P_{Gas}$ and $P_{Air}$ are the transmitted power of light via the gas cell in the presence of gas and in the absence of gas, respectively. The transmitted light is detected by an optical sensor (here, a silicide detector). This detector produces an output signal as current, with the level of current depending on the intensity of irradiation. The relation between incident light intensity and diode current are expressed as in equation 4.

$$I_{ph} = RP \qquad (4)$$

Where $I_{ph}$ and R are the photo-current and responsivity of the detector, respectively, and $P$ is the power of the incident light. Hence, in the presence of gas, the intensity of light reduces due to partial absorption, and consequently, the diode current will decrease.

Considering $I_{Gas} = RP_{Gas}$ and $I_{Air} = RP_{Air}$ as the photocurrent of the detector in the presence and in the absence of the gas, respectively, equation 3 can be written as follows:

$$I_{Gas} = RP_{Air}\exp(-k_{eff}cl) \qquad (5)$$

$$I_{Gas} = I_{Air}\exp(-k_{eff}cl) \qquad (6)$$

$$I_{Gas}/I_{Air} = \exp(-k_{eff}cl) \qquad (7)$$

Therefore, the concentration of gas can be determined by using equation 7. In the same way, the effective absorption coefficient can be determined for known concentrations.

## 2  NiSi/n-Si detector fabrication and characterization

The Schottky diode was fabricated [10] using n-type Si (100) wafer of resistivity 1–10 Ω-cm. Cleaning of wafer was performed to remove native oxide and organic contaminants from the surface. A 100 nm thick $SiO_2$ layer was grown by wet oxidation process for contact pad deposition. Backside of the wafer was etched by Buffered Hydro Fluoric (BHF) acid and subsequently, $n^+$ region was made by ion implantation followed by 30 s Rapid Thermal Annealing (RTA) at 950 °C. Optical lithography was used to carve out a window of $0.5 \times 1\,mm^2$ and selective removal of $SiO_2$ was done from the surface by the BHF. Pattering for top electrode was carried out on the $SiO_2$ for Ni deposition. After patterning, wafer was dipped into BHF to remove native oxide formed during the process. Following the removal of native oxide, the wafer was immediately loaded in physical vapour deposition chamber for Ni deposition. Deposition was performed at a base vacuum of $5 \times 10^{-6}$ mbars. A 10 nm Ni film was deposited on the patterned Si substrate followed by lift-off. Subsequently, RTA was performed at 500 °C for 60 s for silicide formation. The unreacted Ni was removed by treating with an acid mixture ($HNO_3$:HCl = 1:5 for 60 s). Finally, Au was deposited for top contact ($1 \times 1\,mm^2$) and Ti/Au was deposited for back ohmic contact. Thickness of the silicide was measured using Cross-sectional Secondary Electron Microscopy (SEM) (Raith-150) technique. X-ray Photoelectron Spectroscopy (XPS) (PHI5000VersaProbe-II) and Raman spectroscopic measurement (RAMNORHG-2S) were performed to analyse the Ni-Si phases. The area of top silicide contact has been measured using microscope. The electrical characterization of diode was performed using Keithley 4200 instrument. Optical response was measured using Keithley 2400 under illumination of a tungsten lamp with a mono-chromator arrangement.

The Raman spectroscopic analyses indicate that the developed phase of the silicide is NiSi with a characteristic peak at 199, 217, 294, and 363 cm$^{-1}$ [10]. Similarly, XPS spectrum with peaks at 853.9 eV and 871 eV of Ni2p3/2 and Ni2p1/2 corresponds to NiSi phase [10]. From the cross-sectional SEM imaging it is observed that the developed NiSi film is uniform and the thickness is 27 nm. The current–voltage (I–V) characteristics of the developed device, point to Schottky nature of the diode with a barrier height of 0.62 eV corresponding to cut-off wavelength of 2 µm [10]. To investigate the optical response of SBD due to illumination by the 1.65 µm laser, the SBD was illuminated by a focused infrared laser. The response of the detector was characterized at different reverse bias levels by varying the illumination power. The photocurrent was measured by a source measuring unit (Keithley 2400). The emitting area of the diode is approximately 5 µm$^2$. The characteristics of the sensor are shown in Figure *2*.

When the infrared light is incident on the SBD detector, the total output current is the sum of the photo-current ($I_{ph}$) and the background current ($I_{bg}$, dark current), which can be expressed as

$$I_{tot} = I_{ph} + I_{bg} \tag{8}$$

For an optical detector, $I_{ph}$ is the crucial parameter that is closely related to the responsivity ($I_{ph}$/Power of the laser, equation 4). Therefore, $I_{ph}$ of a detector must be improved to obtain better performance. This improvement can be achieved by applying a higher bias and by increasing the irradiation intensity. A similar behavior can be observed in Figure 2, where the reverse bias and irradiation intensity were varied, revealing an increasing value of $I_{ph}$ with higher reverse bias and higher irradiation intensity. This characteristic can be explained as follows: (i) the higher incident optical power results in a higher amount of carrier generation in the device, there by resulting in higher photo current; and (ii) the increase in reverse bias contributes to rapid sweeping out of the photo-carriers, thereby increasing the photocurrent.

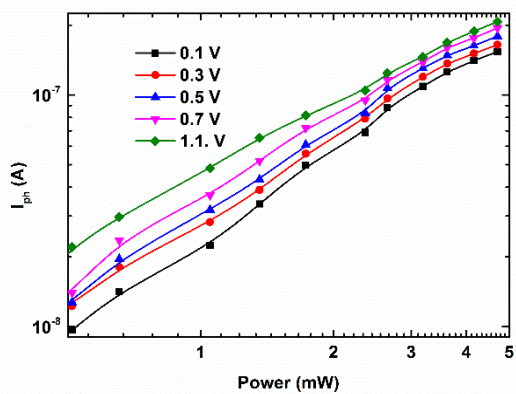

Figure 2: Photocurrent ($I_{ph}$) versus laser illumination power.

The above observation indicates that a higher reverse bias is favorable for detector application. However, this approach comes at the cost of higher dark current, which degrades the signal-to-background ratio and demands careful background signal suppression. To optimize the operating condition of the detector, the dependence of the ratio of $I_{tot}/I_{bg}$ on the reverse bias is studied for different illumination powers (shown in Figure 3). Zero bias is found to be a better choice for detector operation because the signal-to-background ratio is high compared to that under a negative bias condition. Hence, for this experiment of methane sensing, the bias of the detector was kept at zero, and the illumination energy was kept at 2.1 mW. The medium illumination power was chosen to reduce the possibility of heating of the laser LED, which may result in shifting of the emission wavelength.

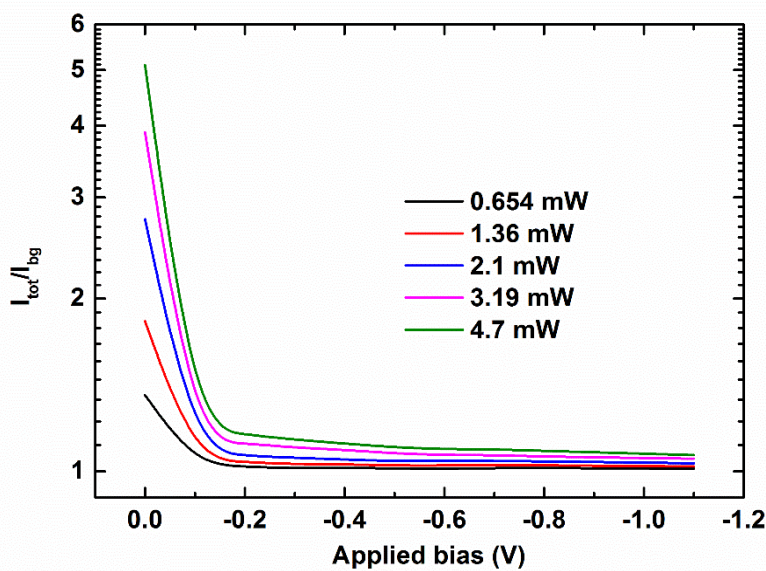

Figure 3: Plot of the ratio of $I_{ph}$ to $I_{bg}$ versus the bias voltage under different illumination powers.

## 3 Design of Set-up and Testing

### 3.1 Setup Design

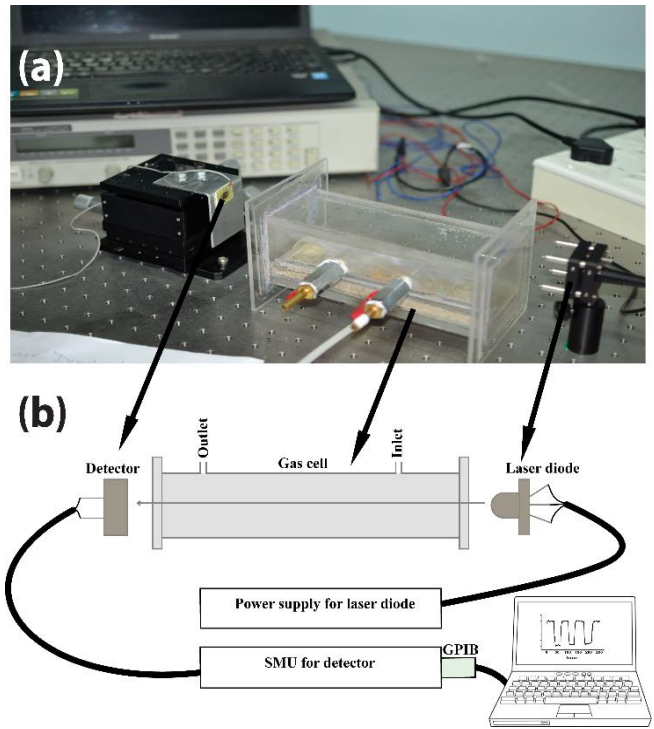

Figure 4: Experimental setup for gas detection: (a) image of experimental setup and (b) schematic diagram of the gas detection setup.

The laboratory setup and schematic diagram of the methane detection system are shown in Figure 4 (a&b). Through the gas inlet, 99.5 % pure methane was allowed to flow through the line of sight of the infrared detector and the source via a gas cell. The gas cell is made of acrylic with quartz plates used as an infrared window. The detection process includes partial absorption of infrared by methane. Thus, as per equation 6, the intensity of light will decrease as it transmits through methane environment, and the reduction in light intensity is detected by the sensor placed on the other side of the source. The experiment was repeated with ambient air (where no absorption is expected) in the gas cell to compare the results. The data acquisition is performed using a source measuring unit (Keithley 2400) along with a GPIB (Keithley 488 A) interface to a computer running the MATLAB program. An Agilent 6645A DC power supply is used to control the input voltage (and thus the output optical power) of the laser diode. The experiments were conducted for various concentrations of methane.

### 3.2 Methane Sensing

As stated in equation 8, the total current of the detector in the presence of gas is

$$I_{tot-Gas} = I_{Gas} + I_{bg} \tag{9}$$

In addition, in the absence of gas (which is ambient air), the total current is

$$I_{tot-Air} = I_{Air} + I_{bg} \tag{10}$$

Thus,

$$I_{Gas}/I_{Air} = (I_{tot-Gas} - I_{bg})/(I_{tot-Air} - I_{bg}) \tag{11}$$

Therefore, combining equations 6 and 11:

$$I_{Gas}/I_{Air} = (I_{tot-Gas} - I_{bg})/(I_{tot-Air} - I_{bg}) = \exp(-k_{eff}nl) \tag{12}$$

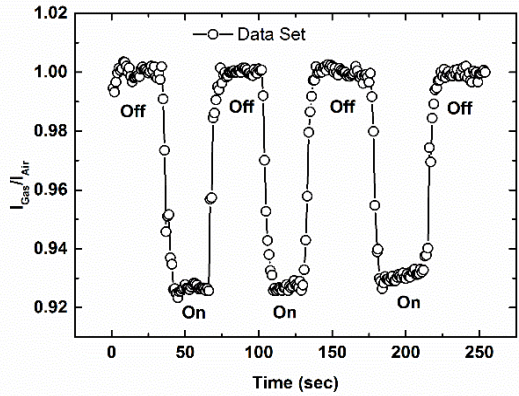

Figure 5: Gas on/off measurement to observe the speed of the response.

As per equation 12, the ratio of $I_{Gas}/I_{Air}$ will decrease when gas is introduced in the gas cell due to the partial absorption of light. The dynamic characteristic of the ratio in the presence and absence of methane in the chamber is shown in the Figure 5. The label 'off' represents the cell without gas, and the label 'on' represents the cell filled with gas. It is evident from figure 5 that that the device is capable of detecting the presence or absence of methane instantaneously.

3.2.1. Response and Recovery time

It is well known that response and recovery are two important quality factors of a gas sensor. The first factor implies how fast the sensor senses the presence of gas, whereas the latter indicates how fast it senses the absence. To estimate these aspects of this sensor, an experiment was performed (Figure 6). The result of this experiment indicates that the device is fast enough to respond to the presence or absence of methane. The response and recovery of the detector is studied and found that they have an exponential dependence with time (as shown in Figure 6). The response ($\tau_d$) and recovery ($\tau_r$) time constants are determined by exponential fitting of the dynamic plots with time, and the values are found to be 2.14 and 2.44 sec, respectively. Similar work was performed by Schmauch (1959)[32]; our result is in close agreement with his work. Our investigation shows that this device provides a quick response and can be deployed in mines for real-time fire hazard mitigation.

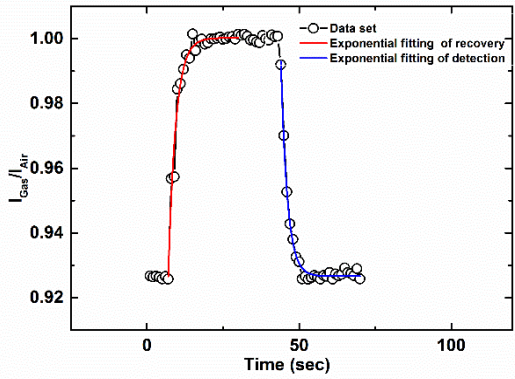

Figure 6: Methane response time of recovery and detection.

3.2.2. Sensor Response and Methane Concentration

To evaluate the sensor response for different concentrations of methane, experiments were performed in which the sensor is exposed to different concentration of methane. Initially, the data acquisition was performed by allowing ambient air into the gas cell. The methane was then purged in the gas cell, where the volume of the methane was controlled by a mass flow regulator. The concentration of gas inside the cell varied from 3% to 9%. The lowest level of detection of gas concentration is up to 3% by volume. The lower explosion level of methane is 5%, which is well above the range of our detection limit. The response characteristic of the device at different concentrations is shown in Figure 7. It is evident from the figure that the $I_{Gas}/I_{Air}$ ratio decreases with the increase in concentration of gas in the system. This result is attributed to the increase in the absorption of light by the gas molecule along the optical path with the increase in gas concentration, followed by the decrease in photo-current. The variation of $I_{Gas}/I_{Air}$ with different methane concentrations in the cell is plotted in Figure 8. The absorption cross-section was calculated and found to be $4.3 \times 10^{-2}$ cm$^2$/mole, which is comparable to that of the observations of Okajima *et al.*[33]. However, it is pertinent to mention here that estimates of Okajima *et al.*[33] are based on the costly and sophisticated component like InGaAs detectors.

The lower detection level of the system (3% in air) can be further improved by using better components and arrangements, such as a high power laser, increasing the detection area (increasing lasing spot size), and multiple path gas detection by mirror arrangement [12,23,24,34]. It is evident from the results that the proposed system can detect methane at lower concentrations (3%) than the explosion level (5%). Hence, this device can detect methane before it reaches the critical explosion level. Measurements of very low concentrations of methane (in terms of ppm) can be possible when this detector is used in a frequency modulated system.

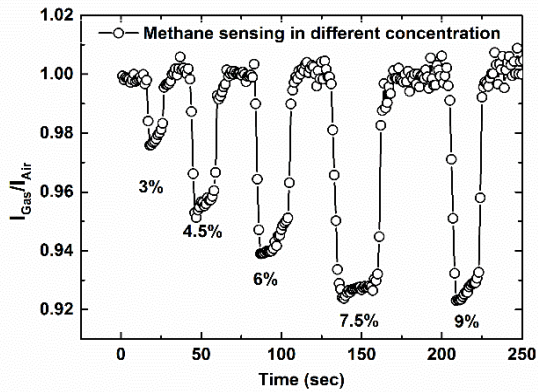

Figure 7: Real-time response of the detector at different concentration.

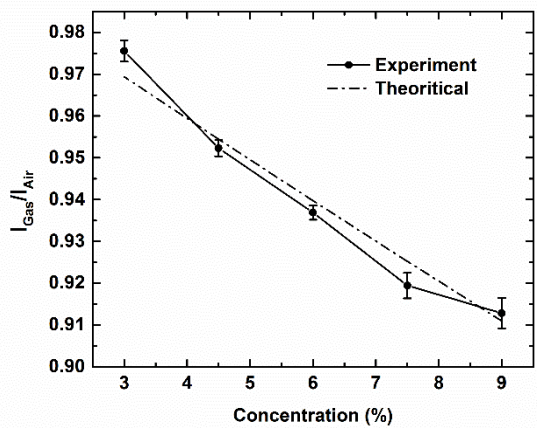

Figure 8: Diode response with various methane concentrations to determine the absorption cross-section.

3.2.3. Hydraulic properties vis-à-vis Methane Diffusion Rates

To understand the suitability of the developed sensor in real time monitoring, methane flux and response time to detect the changes in fluxes were measured. For this purpose, methane was allowed to diffuse through test beds with different porosity and permeability (eg. sand, clayey sand, clay) and methane flux was detected while passing through the medium. The time taken by the gas to diffuse through the medium was recorded by noting down the purging and detection time. List of the detection time vis-a-vis different medium is shown below (Table 1).

Table 1: Table of diffusion time with different gas barrier

| Medium | Relative Gas permeability | Response time |
|---|---|---|
| Open Air | High | 1 sec |
| Sandy clay | High | ~2 sec (1.68 sec) |
| Saturated sand | Low | 6.8 sec |
| Clay | Low | 5.6 sec |

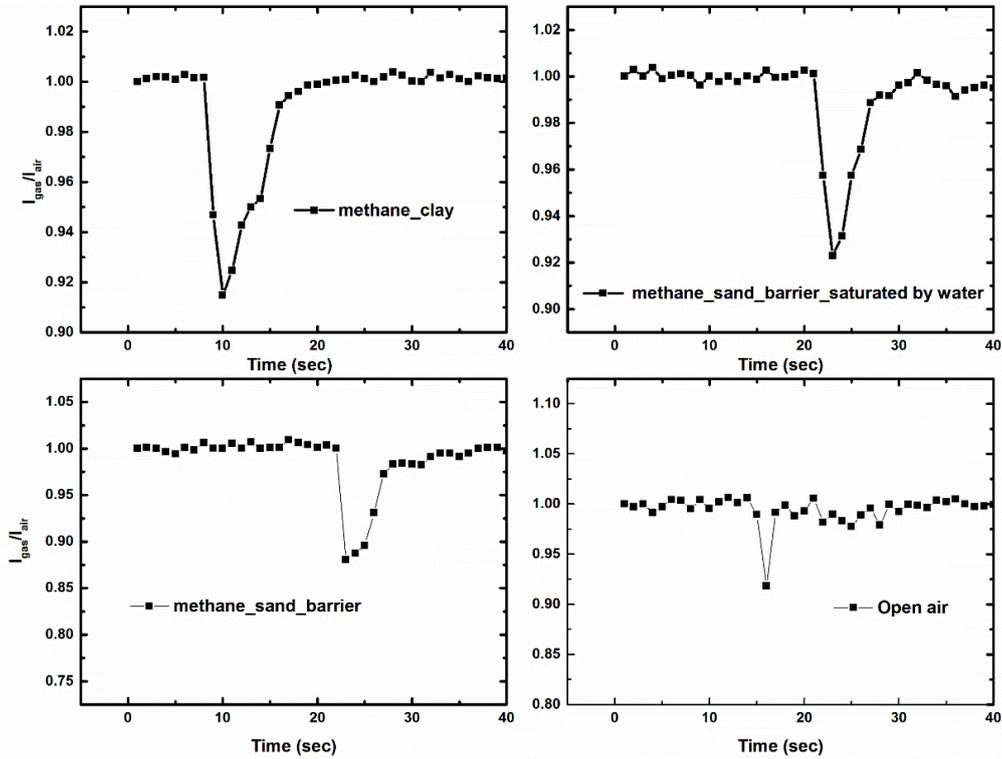

Figure 9: Gas diffusion time with different barrier for methane

It can be observed from the Figure 9 and Table *1* that as the permeability of the medium increases the diffusion time decreases and the flux can be estimated even in the swiftly diffusing sands. This observation indicates that the device can be effectively used in quantifying the emission fluxes in environments like micro seepages and mine venting.

## 4 Conclusion

This work highlights the efficacy of NiSi/n-Si Schottky diode as an optical sensor for methane detection. The barrier height of the SBD is designed and characterized for 1.65 µm (methane absorption feature) infrared radiation. Based on the I-V characteristics and optical response, the sensor is observed to detect the methane concentration as low as 3%. This performance is par excellent for estimating the methane fluxes in bore holes and mine vents. The quick response (~2 sec) of the sensor for the methane buildup is an added advantage in real time operations. Finally, unlike other optical methane detectors, the low cost fabrication route of NiSi/n-Si SBD empowers its massive production and wider deployment in mines, bore holes and other places where methane emission monitoring is needed.

**Acknowledgement:**

The authors would like to acknowledge the financial support by TATA Consultancy Service (project code: 13TCSIRC004) for the financial support. The authors also would like to thanks Mr. Aranb Pattanayak and Mr. Soumitra K. Nayak, Centre for Nanotechnology and Science, IIT Bombay for their kind help during the gas detection.